# Local structure in $Mg_{1-x}Al_xB_2$ system by high resolution neutron diffraction

G. Campi[1*], A. Ricci[2,3] and A. Bianconi[2]

[1]Institute of Crystallography – CNR, Via Salaria Km 29.300, 00015, Monterotondo – Roma, Italy

[2]Department of Physics, Sapienza University of Rome, P. le A. Moro 2, 00185 Rome, Italy

[3]Deutsches Elektronen-Synchrotron DESY, Notkestr. 85, D-22607 Hamburg, Germany

*E-mail: gaetano.campi@ic.cnr.it

**Abstract**

The local structure in high temperature superconductors is nowadays considered a key point for understanding superconductivity mechanism. $MgB_2$ has a well-known simple structure; but its local structure remains quite unexplored. This is due to the fact that typical x-ray local probes, such as EXAFS, fail when used to study local structure of light atoms, such as Mg and B. We used high resolution neutron diffraction with pair distribution function (PDF) analysis for investigating disorder on the atomic scale in the Al doped $Mg_{1-x}Al_xB_2$ system. The results indicate an anisotropic structural inhomogeneity along the *c*-axis that could be related with the delocalized metallic-type bonding between Boron layers.

## 1. Introduction

The discovery of superconductivity at 39 K in the $MgB_2$ binary system [1] has triggered enormous interest in the condensed matter community because of the apparent simplicity of its electronic and lattice structure. It has been soon clear that this binary material made of non magnetic light elements [2] was the practical realization of a high $T_c$ superconducting material by design described in the 1994 patent and papers [3]. In fact, it is a very simple heterostructure at atomic limit: a superlattice of atomic layers where the chemical potential is tuned near two Lifshitz transitions for the appearing a new Fermi surface spot (due to the 2D σ band of $p_{xy}$ orbitals within the boron layers) and the closing of a neck typical for superlattices material near a band edge [4]. The predicted behaviour of the gap ratio in this clear case of two gaps superconductor [5] has been well verified in ten years of research [6] by tuning the chemical potential around the shape resonance for the superconducting gaps. In this scenario the structure of the superlattice of atomic layers play a critical role in the control of the Lifshitz critical points that allow new high $T_c$ superconductors by material design [7].

The complex phase separation scenario approaching the Lifshitz transition critical point has been observed in $Mg_{1-x}Al_xB_2$ by changing the Al content [8-10]. The phase separation shows frustration and multiples competing phases [11] like in the recent discovered Fe-based superconductors [12]. Therefore the $MgB_2$ constitutes a good example of two bands superconductor where, apart from the BCS type intra-band pairing, the inter-band pairing appears to have significant importance in increasing the superconducting critical temperature; furthermore, approaching the Lifshitz transition the system shows also phase separation. Therefore there is a need for the study of the complexity of the material and electronic structure by investigation of the structural properties of the system by changing the carrier concentration by chemical substitution using multiple experimental methods probing the local structure like EXAFS, XANES and the neutron pair distribution function (PDF).

In this work we have used high resolution neutron pair distribution function (PDF) measurements to investigate the atomic scale lattice inhomogeneity of $Mg_{1-x}Al_xB_2$ with variable Al content. In fact, since $Mg_{1-x}Al_xB_2$ is made of light elements with shallow core levels the atomic scale inhomogeneity cannot be investigated by EXAFS using hard x-ray photo-absorption by deep core levels.

## 2. Experimental results and discussion

Time-of-flight neutron powder diffraction data were measured on the NPDF diffractometer at the Manual Lujan, jr., Neutron Scattering Center (LANSCE) at Los Alamos National Laboratories. The powdered $Mg_{1-x}Al_xB_2$ were sealed inside extruded cylindrical vanadium containers. An empty container mounted and the empty instrument were all measured, allowing us to asses and subtract the instrumental background. The scattering from a Vanadium rod was also measured to allow the data to be normalized for the incident spectrum and detector efficiencies. Standard data corrections were carried out using the program PDFGETN [14], providing the total scattering structure function, S(Q). This was converted to the PDF, G(r), by a Sine Fourier transform [15]. We have collected diffraction pattern up to a high momentum transfer Q=40 Å$^{-1}$. The PDF patterns have been fitted over the range 1-18 Å using a hexagonal crystal structure P6/mmm. Lattice parameters and scale factor, with all atoms constrained to move in an ellipsoid having the same planar axis and centred on their average positions, were allowed to vary [16]. The measurements have been carried out on several $Mg_{1-x}Al_xB_2$ samples with x=0, 1/12, 2/12, 3/12, 4/12, 5/12, 6/12, at room temperature (T~300K).

We found the same $\sigma^2(Mg)=\sigma^2(Al)$ for Mg and Al atoms in the doped $Mg_{1-x}Al_xB_2$ samples. The reduced structure function Q[S(Q)-1] for the pure $MgB_2$ at T=300 K is shown in Fig. 1. The

Bragg peaks are clearly visible up to 25 Å$^{-1}$ and reflect both the long range order of the crystalline samples and the small amount of positional disorder on the atomic scale. Instead, the oscillating diffuse scattering in the high-Q region, shown in the insets of Fig. 1, contains mainly local structural information. The corresponding reduced PDF, G(r), is shown in Fig. 2. The features of the NPDF diffractometer, combined with the described experimental setup allowed us to obtain high quality PDFs, as can be noted looking at the modeled fits (full line) of the G(r)s in Fig. 2.

The reduced structure functions and the PDF's were carried out for all seven $Mg_{1-x}Al_xMgB_2$ samples studied in this work. The local structure properties have been extracted from the analysis of the PDFs peaks and compared with the long range average structure, obtained by the refinement of the crystallographic parameters. The position of each of the first four peaks in the G(r), named R1, R2, R3, R4, determined by fitting Gaussians, gives the first four interatomic distances in the samples. Being Mg, B(1) and B(2) the atoms in the unit cell of $MgB_2$, R1 and R3 correspond to the in-plane B(1)-B(2) first neighbours distances and the a-axis respectively; R2 is the Mg-B(1)=Mg-B(2) bond length, connecting Mg(Al) and B first neighbour atoms on adjacent layers, while the 4$^{th}$ peak, R4, has two component: the first one is the c-axis and the second one represents the distance between B(1) and B(2) in two different and adjacent Boron planes.

Our main finding come out from the comparison of the average interatomic distances corresponding to the first 4 peaks obtained from the structural refinement, and the interatomic distances obtained from the position of the peaks in the G(r) curves. There is a good agreement between the interatomic distances obtained by two analysis described above, except for the c-axis, contributing to the 4th peak R4. This is shown in Fig. 3, where we have plotted the average *c*-axis obtained from the structural refinement (full line) and the positions of the PDF peaks corresponding to the first four *c*-axis multiple divided by their own multiplicity; a clear discrepancy between the structural refinement data (full line) and the positions of the first three RDF peaks corresponding to the first three *c*-axis multiple is evident. This inconsistency disappears at 4c ~ 14Å interatomic distance where the decreasing *c*-axis with Al concentration is in agreement with the average refined values indicating a strong dependence from misfit strain between layer as has been found in all high T$_c$ superconductors [17-19]. Indeed, Al substitution induces a tensile strain in the structure favouring a larger elasticity along c-axis. The decreasing of the *c*-axis is also shown in Fig. 4, where the PDF peaks corresponding to the R4 peak (*c*-axis) (left panel) and to *4c*-axis (right panel) for each sample are reported. This indicates that the local structure of this system deviates from the average one along the *c*-axis in the interatomic distance range of about 14 Å. This deviation could be ascribed to a significant local lattice disorder along the *c*-axis due to variable distances between adjacent Boron planes; this could be related with the delocalized $p_{xy}$ orbitals between Boron layers.

### 3. Conclusions

In summary, we showed the local structure properties of $Mg_{1-x}Al_xB_2$ system using atomic pair distribution functions (PDF's) obtained from high resolution neutron powder diffraction measurements. By comparing the local structure with the average one, we have found significant deviations from the average structure on the interatomic distances within about 14 Å indicating local structural inhomogeneity. The present results can be compared with the non trivial complexity of multiscale phase separation observed in cuprates. These materials show phase separation at atomic scale as probed by EXAFS [20] related with the modulation of the orbital symmetry [21, 22]. The phase separation is associated with polaronic carriers [23] in a narrow band that coexist with free carriers in a wide band [24, 25]. The phase separation is observed at the nanoscale with the coexistence of pseudo-gap electronic matter with superconducting matter and it extend to the mesoscale to micron scale as observed by scanning x-ray microdiffracion [26, 27]. This has been confirmed by time resolved x-ray diffraction experiment probing the self organization of dopants in the annealing process [28]. Recently the investigation of the iron based superconductors has provided evidence of lattice complexity [29] being in the proximity of a structural phase transition [30] and a Lifshitz critical point [31]. In conclusion we have provided evidence for local lattice distortion in doped magnesium diboride that point toward a complex inhomogeneous structure favoring high Tc as it has been recently proposed [32].

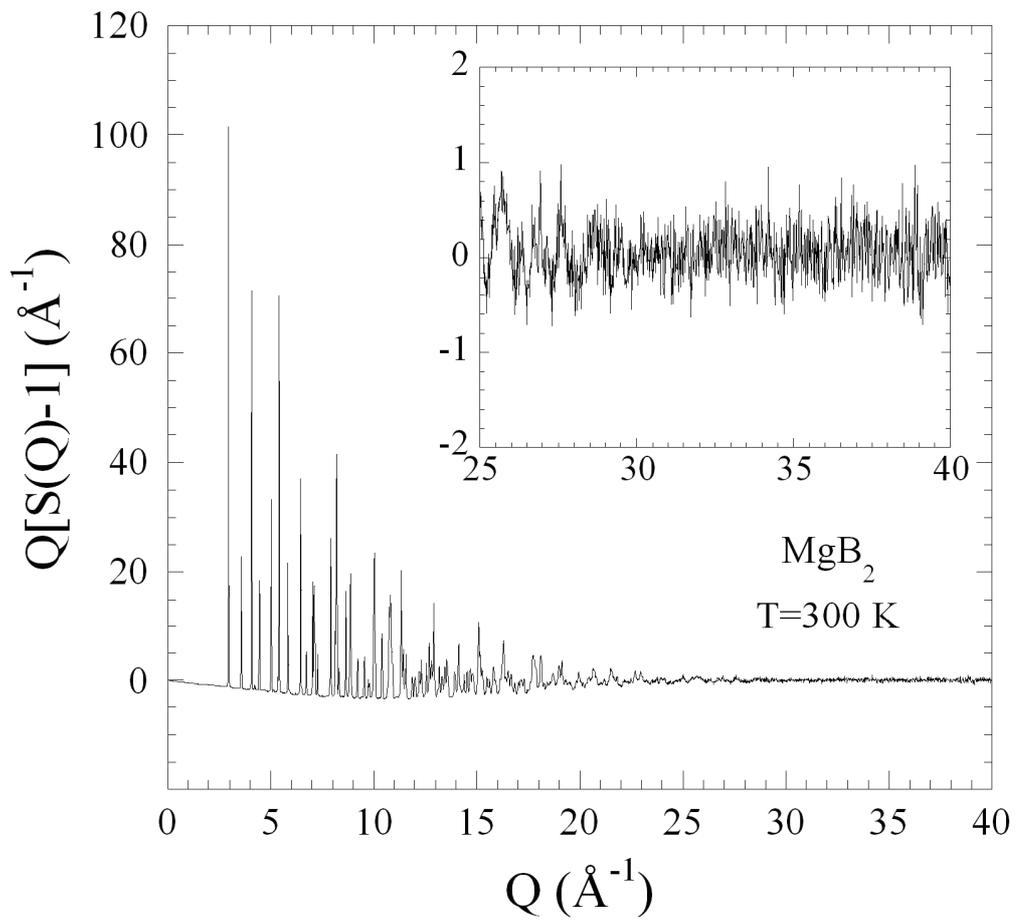

**Fig. 1** The reduced structure function *Q[S(Q) – 1]* for MgB$_2$ measured at 300 K

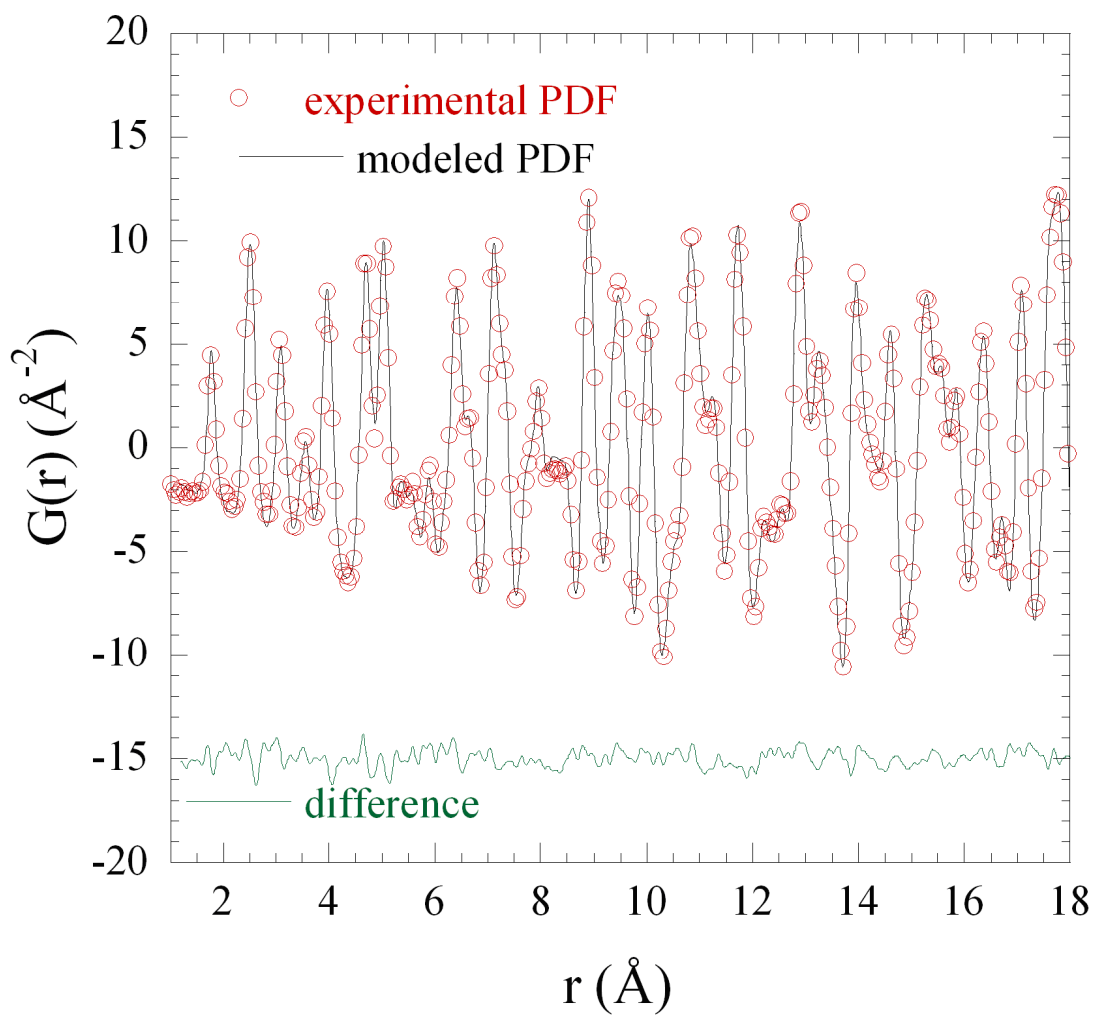

**Fig. 2** The experimental PDF $G(r)=4\pi r[\rho(r) - \rho_0]$ for $MgB_2$ measured at 300 K (circles); the line is the average structure refinement obtained by a least-squares approach.

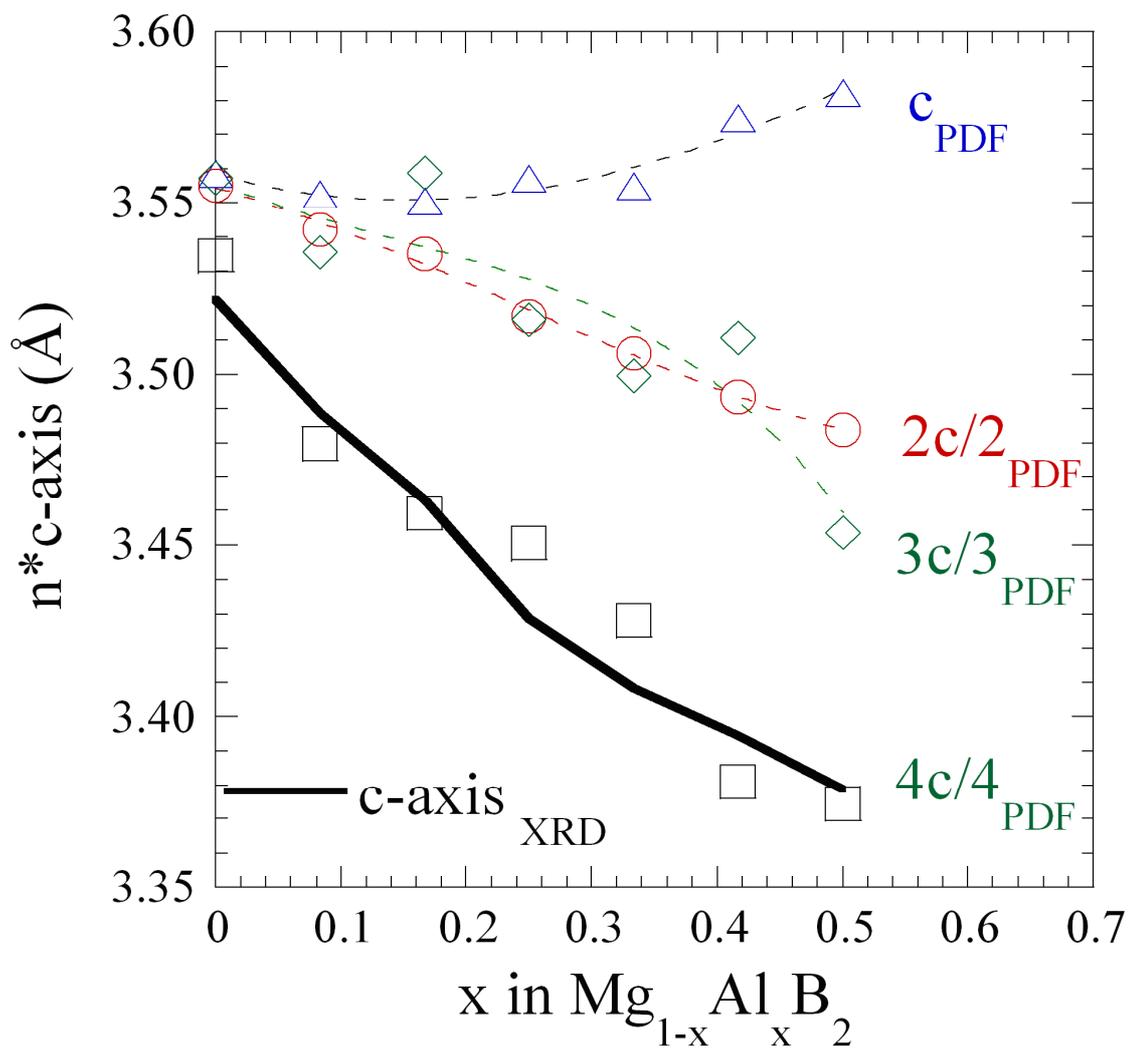

**Fig. 3** Position of the four peaks in R(r) functions, named $c_{PDF}$, $2c/2_{PDF}$, $3c/3_{PDF}$, $4c/4_{PDF}$, corresponding to the first four $c$-axis multiple, each of which has been divided by its multiplicity; the position of the peaks have been determined by fitting Gaussians. The full line indicates the $c$-axis value obtained from the structural refinement.

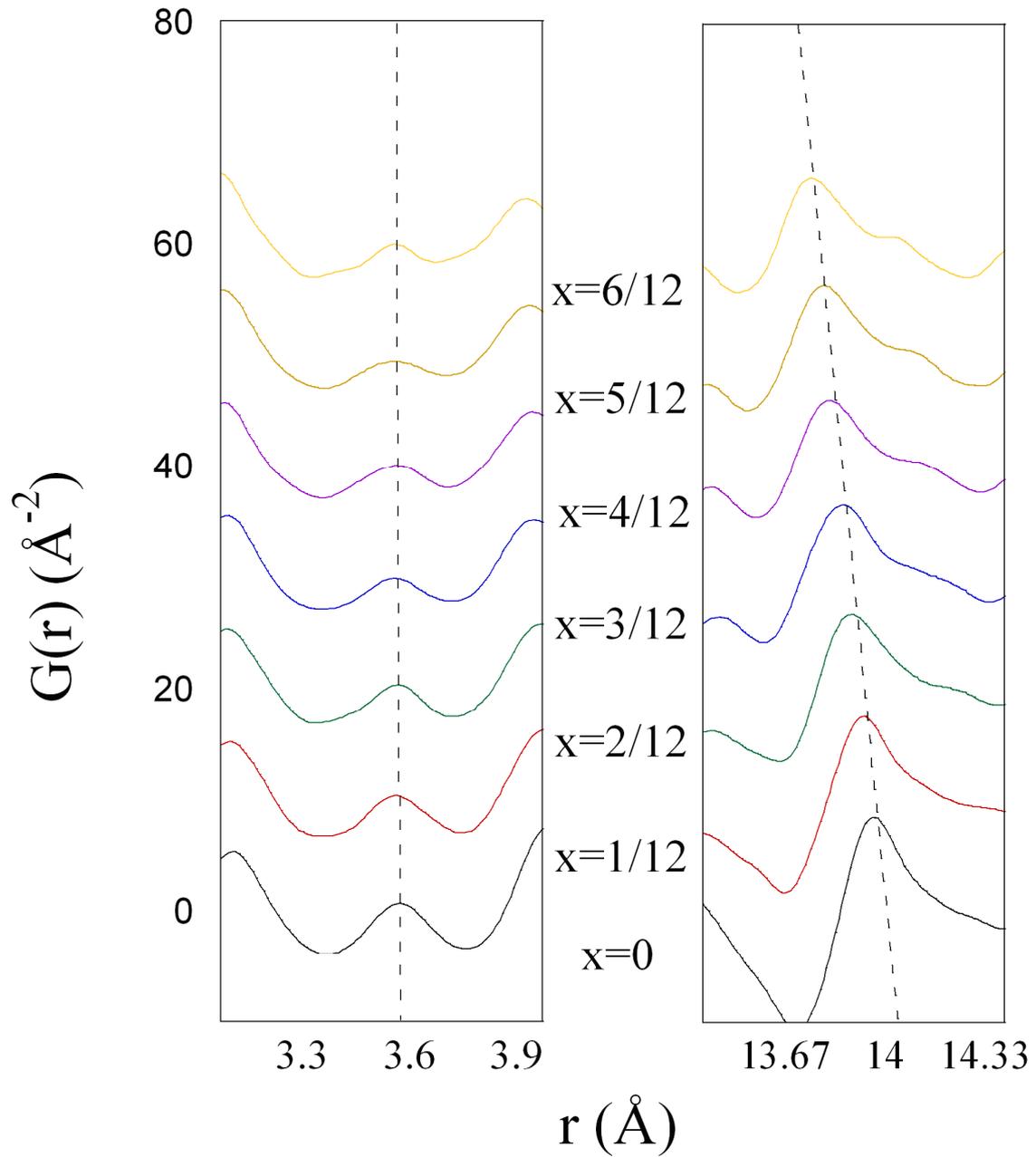

**Fig. 4** PDF peaks corresponding to the *c*-axis (left panel) and to *4\*c*-axis (right panel) for different samples; here, at the interatomic distance R~14Å~4c, the contraction of the *c*-axis measured in the PDF spectra gets equal to the structural refined values for the c-axis.